# Effect of (Cu/Fe)O$_5$ bipyramid size and separation on magnetic and dielectric properties of rare earth layered perovskite LaBaCuFeO$_5$ and LuBaCuFeO$_5$

Surender Lal, C. S. Yadav and K. Mukherjee

School of Basic Sciences, Indian Institute of Technology Mandi, Mandi-175005(H.P.), India

**Abstract**

We report structural, magnetic and dielectric properties of layered perovskite materials LnBaCuFeO$_5$ (Ln = La and Lu). LaBaCuFeO$_5$ shows magnetic cluster glass behavior below 60 K owing to the competing ferromagnetic and antiferromagnetic exchange interactions. Glassy dynamics of electric dipoles has also been observed in the vicinity of the magnetic glass transition temperature. The presence of significant coupling between spin and polar degrees of freedom results in the multiglass feature in LaBaCuFeO$_5$. The LuBaCuFeO$_5$ compound undergoes YBaCuFeO$_5$ like commensurate to incommensurate antiferromagnetic transition at 175 K. Large magnetic irreversibility below 17 K in this compound suggests the presence of strong spin anisotropy. In addition, in this compound the interaction between the dipoles is not strong enough, which results in the absence of glassy dynamics of electric dipoles. The contrasting behavior of two compounds is possibly due to variation in the ferromagnetic and antiferromagnetic interactions along c-axis, which is the manifestation of structural modification arising out of the difference in the ionic radii of La and Lu.

Keywords: magnetic properties, cluster spin glass, glassy dynamics, multiglass, dielectric relaxations.

## I. Introduction

Multiferroics materials where one ferroic order is controlled by another ferroic order, are relatively rare as it is difficult to have two ferroic orders in a single crystallographic structure [1]. The weak coupling between magnetic and electric dipoles limits their usage for practical device applications. However, the research in multiferroics materials attracted immense attention when spin driven ferroelectricity was found in magnetic materials [2–4]. In such systems, magnetic and dielectric transitions are in close vicinity of each other and electric polarization can be switched by magnetic field. Below the magnetic ordering temperature, the time reversal symmetry is broken whereas the spatial inversion symmetry is broken in case of ferroelectric system. In case of multiferroics both time reversal and spatial inversion symmetries are broken, which lead to the finite magneto-electric coupling between the magnetic and electric order parameters [5,6]. Spatial inversion symmetry can also be broken due to canted or spiral antiferromagnetic (AFM) ordering that leads to anti-symmetric exchange interactions. It is well known that Dzyaloshinskii-Moriya interaction results in the breaking of spatial inversion symmetry at low temperature due to the spiral magnetic structure, exchange-striction mechanism or spin-direction dependent metal-ligand hybridization [7–11]. These mechanisms are commonly responsible for the observation of ferroelectricity in multiferroics.

In this context, $YBaCuFeO_5$ compound has been extensively investigated for multiferroic properties due to close proximity of magnetic and dielectric transitions around 200 K [12]. In view of the novel features exhibited by $YBaCuFeO_5$ [12–15], it is of interest to study the properties of other non-magnetic rare-earth analogues of this layered perovskite compound. As magnetic and dielectric properties of these compounds are strongly dependent on structure, it is expected that non-magnetic rare-earth ions with different ionic radii might give rise to unique physical properties in the respective compounds.

In this manuscript, we report the structural, magnetic and dielectric properties of $LaBaCuFeO_5$ and $LuBaCuFeO_5$ compounds. The La and Lu elements lie at two extreme ends of the 4$f$ block of the periodic table. The $La^{3+}$ has the highest ionic radii (1.032 Å) whereas $Lu^{3+}$ has the minimum ionic radii (0.861 Å) among all the 4$f$ block elements. Moreover, both of these rare-earth elements are non-magnetic in their +3 oxidation state, which indicate the absence of additional magnetic contribution in these compounds similar to $YBaCuFeO_5$. The study of these two compounds will help in identifying the influence of structure on the physical properties that are observed for YBaCuFeO$_5$ [14]. We found that LaBaCuFeO$_5$ shows magnetic cluster glass behavior along with glassy dynamics of the electric dipoles below 60 K. On the other hand, the AFM ordering persists in LuBaCuFeO$_5$, along with the presence of strong spin anisotropy below 17 K. We observe that the physical properties of these rare-earth layered perovskite compounds are strongly dependent on structure and the expansion/compression of the unit cell volume and shows contrasting behavior.

## II. Experimental details

The compounds LaBaCuFeO$_5$ and LuBaCuFeO$_5$ are prepared under similar conditions as mentioned in our previous work [16–18]. Powder x-ray diffraction at room temperature is performed using a Rigaku smart lab diffractometer using CuK$\alpha_1$ radiation. DC and AC magnetic susceptibility measurements are carried out with a SQUID magnetometer (Quantum Design USA). Dielectric measurements as a function of temperature and magnetic field are performed using a Hioki LCR meter with a sample insert from Cryonano Lab integrated with PPMS. Measurements are made on a thin parallel plate capacitor with Cu wire. Contacts are made using silver paste.

## III. Results
### i). Structural properties of LaBaCuFeO$_5$ and LuBaCuFeO$_5$

Rietveld refined x-ray diffraction patterns of LaBaCuFeO$_5$ and LuBaCuFeO$_5$ are shown in the figure 1. These compounds are formed in the tetragonal structure with P4mm space group [19–22]. The parameters obtained from the fitting are tabulated in Table 1. It is noted that the lattice parameters decrease from La to Lu, in accordance with the lanthanide contraction rule. The structure of LnBaCuFeO$_5$ (Ln = Rare-earths) can be described as consisting of [CuFeO$_5$]$_\infty$ double layers of the corner sharing CuO$_5$ and FeO$_5$ pyramids along the *c*-axis, containing Ba$^{2+}$ ions [19,23,24]. These [CuFeO$_5$]$_\infty$ double layers are separated by Ln$^{3+}$ planes. Here we would like to mention that thermo-gravimetric analysis (TGA) performed for the GdBaCuFeO$_5$, HoBaCuFeO$_5$ and YbBaCuFeO$_5$ compounds (belonging to the same family) ruled out the non-stoichiometry of oxygen [18]. Since the method of preparation, for the current studied compounds is similar to that of Ref.[18], we expect stoichiometric oxygen content in these compounds. The structure of LnBaCuFeO$_5$ can be described as consisting of [CuFeO$_5$]$_\infty$ double layers of the corner sharing CuO$_5$ and FeO$_5$ pyramids along *c*- direction, containing the $Ba^{2+}$ ions [19,23,24]. These $[CuFeO_5]_\infty$ double layers are separated by $Ln^{3+}$ planes. The $CuO_5/FeO_5$ forms the bipyramids, which are linked together via apical oxygen and are aligned ferromagnetically within the bipyramids. However, these bipyramids are aligned antiferromagnetically along the *c*-direction, giving rise to spiral antiferromagnetic magnetic ordering in this direction [14]. The change in the structure due to differences in the ionic radii of the rare-earth ions may have an influence on the magnetic and dielectric properties of the respective compounds[18]. As observed from the table 1, the inter-pyramidal distance increases whereas the distance between the magnetic ions within the bipyramids decreases for $LaBaCuFeO_5$ in comparison to $YBaCuFeO_5$. In case of $LuBaCuFeO_5$, the Cu/Fe inter-pyramidal distance decreases but distance within the bipyramids is comparable to that of $YBaCuFeO_5$. It is to be noted that in case of $LaBaCuFeO_5$, the bond distance of Cu/Fe from the apical oxygen decreases as compared to $YBaCuFeO_5$ whereas for $LuBaCuFeO_5$, this bond distance increases.

## ii). Magnetic Properties of $LaBaCuFeO_5$

Figure 2(a) shows the temperature dependent DC susceptibility of $LaBaCuFeO_5$ under zero field cooled (ZFC) and field cooled (FC) conditions at 100 Oe. The ZFC curve shows a broad peak around 60 K and a strong bifurcation between the ZFC and FC curves is observed below 60 K[25]. Such feature is observed in various systems due to the presence of a glassy magnetic state, canted magnetic state or in anisotropic ferromagnets [26]. The isothermal magnetization at 2 and 300 K is shown in the inset of the figure 2(a). The curve at 2 K shows a weak magnetic hysteresis. However, both the curves do not saturate even at the highest measuring field of 7 T. In order to identify the magnetic state below 60 K, the ZFC and FC magnetization were measured at different magnetic field in temperature range of 2 to 120 K as shown in figure 2(b). The bifurcation persists even in presence of the magnetic field of 30 kOe. However, the peak temperature of the ZFC curve shifts towards lower temperature with an increase in magnetic field. This type of feature is generally observed in compounds that exhibit glassy spin dynamics[27]. In order to substantiate whether the glassy spin dynamics is present in this compound, the peak temperature ($T_p$) of ZFC curve were analyzed with the following temperature-field phase transition relation equation[28]

$$T_P(H) = T_g(0)\left(1 - AH^p\right) \quad\ldots\ldots (1)$$

where $T_g(0)$ is the transition temperature at zero magnetic field, parameter $A$ depends on the anisotropy strength and the value of exponent $p$ depends on the strength of the magnetic anisotropy relative to applied magnetic field. Generally, for a magnetic glassy system, the shifting of the peak temperature with magnetic field follows de Almeida and Thouless (AT) line or Gabay and Toulouse (GT) line. The AT line, which usually occurs in Ising spin glasses and is given as $T_p(H) \sim H^{2/3}$, influence the onset of freezing of the spin component longitudinal to field [29]. In contrast, the GT line is usually seen for Heisenberg spin glasses and varies as $T_p(H) \sim H^2$, which predicts the freezing of transverse spin component[30]. The value of the exponent of magnetic field describes two regions, one with strong anisotropy and other with the weak anisotropy. The inset of figure 2(b) shows variation of the peak temperature with magnetic field. The obtained parameters from the fitting of data by the equation 1, are $T_g(0) = 59\pm2$ K, $A = (3.08\pm2.2)\times10^{-3}$ Oe$^{-0.49}$ and $p = 0.49 \pm 0.06$. The obtained value of exponent ($p = 0.49 \pm 0.06$) is lower than $p \sim 0.66$ for the $A$–$T$ line. The deviation from the $A$-$T$ line is due to strong spin anisotropy, and indicates the clustering of the spins. This gives an indication of the presence of glassy phase of the compound. Additionally, the continuous increase in FC magnetization on decreasing temperature (below the peak temperature) also support the presence of glassy magnetic phase [31,32].

In order to further investigate the exact nature of the glassy magnetic phase of the compound, temperature dependence of the AC susceptibility as shown in figure 2 (c) is measured at different frequencies in the temperature range of 2 K to 120 K. For 13 Hz, the curve shows peak around 60 K (similar to that observed from DC magnetization) which shifts towards higher temperature on increasing frequency. This is a common feature of the glassy magnetic state and indicates the absence of long range magnetic ordering in this compound [32–35]. Here we assign this temperature as the freezing temperature ($T_f$), as generally noted for glassy systems. A parameter $\delta T_f$ (called Mydosh Parameter) can be determined by relative shift in $T_f$ per decade of frequency [34]

$$\delta T_f = \frac{\Delta T_f}{T_f \Delta \log_{10} f} \quad \ldots (2)$$

where $\Delta T_f$ is the relative change in the freezing temperature. This parameter is used to classify various glassy magnetic systems. For the spin glass system, the value of $\delta T_f$ is in the range of 0.001 to 0.01[26], whereas, for super paramagnet it lies in the range of 0.3 to 0.5 [34], and for cluster glasses, the value of $\delta T_f$ lies in between to that for spin glass and super paramagnet [36,37]. For the present system the value to $\delta T_f$ have been observed in the order of order of 0.017 which is in analogy with the compounds exhibiting cluster glass behavior [38–41]. The observed freezing is further quantified with Vogel-Fulcher law [42,43], as given below:

$$\tau = \tau_0 \exp\left[\frac{E_a}{K_B(T_f - T_0)}\right] \quad \ldots (3)$$

where $T_0$ is the measure of the interaction strength between the magnetic entities and $E_a$ is the average activation energy. Figure 2 (d) shows the freezing temperature plotted with $\ln\tau$, fitted with equation (3). The obtained values of $\tau_0$ and $T_0$ are ~ $1.5\times10^{-5}$ sec and 57 K respectively. As per literature report, when $T_0 < E_a/k_B$, it indicates to weak coupling while $T_0 > E_a/k_B$ is a sign of strong coupling between the spins[44]. In our case, the value of $E_a/k_B$ is obtained to be 16.8 K, implying a presence of strong coupling. To further substantiate the presence of the magnetic cluster glass behavior, the temperature dependence of relaxation time $\tau$ is analyzed by the power law of the form [34]:

$$\tau = \tau_0 \left(\frac{T_f}{T_g} - 1\right)^{-z\nu} \quad \ldots (4)$$

where $z$ is the dynamic exponent which describes the behavior of the relaxation time, $\nu$ is the critical exponent which describes growth of the spin-correlation length, $T_g$ is true glass transition temperature and $\tau_0$ is the relaxation time of single spin flip. The inset of figure 2(c) shows scaling of $\log\tau$ with the reduced temperature $\xi = [T_f/T_g - 1]$. The red line is fit to the experimental data. The parameters $\tau_0$ and $T_g$ obtained from the fitting are ~ $7\times10^{-7}$ sec and 59.7 K respectively whereas the value of $z\nu$ is ~ 2.2. Thus a slower relaxation is noted (as compared to conventional spin glass) and the value $z\nu$ is similar to that observed for other cluster glass system [40,45–47]. The slower relaxation time implies the presence of cluster of spins, rather than individual spins that undergoes freezing and this compound shows magnetic cluster glass behavior at low temperatures.

The glassy magnetic state of the compound can be characterized by the magnetic relaxation measurements also. Hence, isothermal remnant magnetization measurements have been performed at different temperatures (below the spin freezing temperature) under the following protocols: the compound is cooled in ZFC condition to the measurement temperature and the magnetic field of 100 Oe switched on for 20 minutes. After 20 minutes, the field was switched off and the time evolution of magnetization is measured. The results are presented in figure 3. For the compounds with long-range order, magnetization does not change with time. Our results show that the magnetization decays with time, which indicates the presence of a glassy magnetic phase. The obtained graphs are fitted with the following equation [48]

$$M(t) = M(0) - S \ln\left(1 + \frac{t}{t_0}\right) \dots (5)$$

where $M(0)$ is the intrinsic magnetization and $S$ is the magnetic viscosity [33]. This observed logarithmic relaxation behavior implies that the distribution of energy barrier in this compound is due to presence of groups of spins of various sizes and suggests the existence of collection of relaxation behavior. This observation also supports the presence of cluster glass phase in this compound.

### iii). Dielectric properties of LaBaCuFeO$_5$

Figure 4 (a-b) shows temperature dependence of real (ε') and imaginary (ε") parts of dielectric constant for LaBaCuFeO$_5$ in the temperature range of 10 to 300 K at selected frequencies in an AC bias of 1 Volt. There is more than two order of change in ε' at low temperature below 20 K, which is accompanied by peak shape anomaly in ε". As shown in the inset (i) of figure 4(a), this anomaly shows frequency dispersion in dε'/dT. The peak temperature obtained from the latter curve is fitted with power law relation of equation (4) to understand the electrical dipole relaxation behavior (inset (i) of figure 4(b). Obtained values of relaxation time $\tau_0$ and exponent zν were $4.16 \times 10^{-7}$ sec and 4.42 respectively. It is important to mention here that a similar feature has been reported for YBaCuFeO$_5$ [12,16,49], with the comparable values of $\tau_0$ and zν and corresponds to the formation of dipolar glass state at low temperature in these compounds.

Further increase in temperature, leads to another frequency dependent broad shoulder like anomaly in ε' in the temperature range 50-120 K. This anomaly is evident in dε'/d$T$ plots (shown in inset (ii) of figure 4(a)), as frequency dispersive peaks. The peak shape anomaly is observed at 78 K for $f = $ 10 kHz, which shifts towards higher temperature for higher frequencies. This dielectric anomaly lies close to the magnetic spin-freezing region (See figure 2a). The peak temperature $T_P$ obtained from this curve is analyzed by Arrhenius law and Vogel Fulcher (VF) law [50]. We obtained a better fit with VF law in comparison to Arrhenius law (Inset (ii) of figure 4(b)). The parameters $\tau_0$ and $T_0$ obtained from the fitting are 6.6 x $10^{-11}$ sec and $T_0$ = 64.2 K respectively and the activation energy ($E_a$) is ~0.15 eV. The obtained value of $\tau_0$ indicates the presence of glassy electric-dipole dynamics. The frequency dependencies in both AC susceptibility and complex dielectric support the presence of multiglass behavior in LaBaCuFeO$_5$. The observation of such features is not uncommon and have been reported for other compounds such as $K_{0.989}Li_{0.011}TaO_3$, $Fe_2TiO_5$, $La_2NiMnO_6$, $YbFe_2O_4$ etc.[51–54]. In order to address the possibility of magneto-dielectric coupling, the dielectric constant is measured as a function of the temperature in the presence of magnetic field of 10 kOe. It is observed that dielectric constant shows a small deviation from zero field dielectric constant indicating the influence of the magnetic field on dielectric constant and the possibility of a weak magneto-dielectric coupling (MDE) in this compound. To study the MDE, the magnetic field response of the $\varepsilon'$ was measured in the temperature range of 30 - 100 K. The MDE is expressed as MDE = $\Delta\varepsilon$ (%) = [$\varepsilon'(H)$-$\varepsilon'(0)$]/$\varepsilon'(0)\times100$, where $\varepsilon'(H)$ and $\varepsilon'(0)$ are dielectric constant in presence and absence of magnetic field respectively. Figure 4(c) shows isothermal magnetic field dependence of MDE at $f$ =100 kHz. A weak MDE coupling is observed in the glassy phase of the compound [52,53,55]. It is observed that the MDE is insignificant above 100 K, implying the absence of the coupling above the glassy phase.

### iv). Magnetic Properties of LuBaCuFeO$_5$

Figure 5(a) shows the temperature dependent DC susceptibility of LuBaCuFeO$_5$ under ZFC and FC conditions at 100 Oe in the temperature range of 2 to 390 K. LuBaCuFeO5 is reported to show paramagnetic to antiferromagnetic transition below 485 K [56]. On further lowering of temperature, magnetization shows a clear hump around 175 K, which is in analogy to the commensurate AFM to incommensurate AFM transition of YBaCuFeO$_5$ observed around 200 K[12–14,20,57]. This observed commensurate to incommensurate phase is the manifestation of the crossover from collinear AFM to spiral AFM ordering [13,14]. In LuBaCuFeO$_5$, ZFC and FC curves overlap above 17 K, and sharp bifurcation is observed at low temperatures. This behavior persists even at high field of 10 kOe, albeit with the reduced magnitude at higher field (figure 5 (b)). The peak temperature of ZFC curve does not shift to either side even at high magnetic field of 10 kOe. The independent nature of peak of ZFC curve with respect to the magnetic field indicates the absence of spin glass behavior in LuBaCuFeO$_5$. The AFM ordering persists down to low temperature and strong magnetic irreversibility is observed at 17 K. The isothermal magnetization shows straight-line behavior with the absence of hysteresis with non-saturating behavior, which indicates the AFM nature of compounds and excludes the possibility of the spin flop like transition. To further substantiate the preceding statement, AC susceptibility was measured at different frequencies as a function of temperature (figure 5(c)). Hence, this observed bifurcation below 17 K can be ascribed to the presence of strong spin anisotropy [22]. To determine the exact nature of this low temperature magnetic phase, temperature dependent neutron diffraction is warranted.

**v). Dielectric Properties of LuBaCuFeO$_5$**

Figure 6(a) shows the temperature dependence of ε' for LuBaCuFeO$_5$ at selected frequencies. It is to be noted that the temperature response of ε' shows no anomaly around 17 K. However, a broad shoulder in the curve is observed above 40 K. The corresponding temperature dependence of ε'' is represented in figure 6 (b). The curves are featureless at low temperature; however, a broad peak is observed around 60 K for the curve at 10 kHz. The temperature of this peak is seen to increase with frequency; giving rise to the possibility of the presence of glassy dynamics of the electric dipoles. This variation of the peak temperature in the imaginary part of the dielectric constant is analyzed by Arrhenius and VF law. In contrast to LaBaCuFeO$_5$, a better fit to the peak temperature is obtained by Arrhenius law in this compound[58]. This law is mathematically expressed as $\tau = \tau_0 \exp(E_a/k_B T)$ where $\tau_0$ is the pre-exponential factor, $E_a$ denotes the activation energy and $k_B$ is the Boltzmann constant. The fitting of the peak temperature by Arrhenius law is shown in the inset of figure 6 (b). From the fitting we obtained $E_a$= 0.058 eV and the $\tau_0$ =9.6×10$^{-10}$ sec. The interaction between the dipoles is not enough for the collective freezing of the dipoles. Further, the dielectric constants measured as a function of the temperature in 0 and 10 kOe fields overlap each other, implying the absence of MDE coupling, unlike LaBaCuFeO$_5$. This statement is also substantiated from the fact that isothermal magnetic field dependence of MDE at 10 kHz at 50 and 125 K did not give any significant value as shown in fig 6 (c).

**IV. Discussion**

As reported in Ref.[16], in YBaCuFeO$_5$ the magnetic interactions along *a, b* directions are AFM whereas along *c* direction, it is alternatively ferromagnetic (FM) and AFM. The interaction along *c* direction is the weakest and depends upon distance between the magnetic ions that influences the interactions within the bipyramids and/or inter-pyramids[16]. The LaBaCuFeO$_5$ and LuBaCuFeO$_5$ are the layered perovskite with tetragonal structure and are isostructural with YBaCuFeO$_5$. LaBaCuFeO$_5$ has the higher unit cell volume, whereas for LuBaCuFeO$_5$, the unit cell volume is lower in comparison to YBaCuFeO$_5$. As noted from table I, for LaBaCuFeO$_5$ the lattice parameter is increased, resulting in the decreases in the distance between the Cu/Fe ions within the bipyramid, thereby strengthening the FM interaction. Also, the distance between the bipyramids increases, thereby weakening of AFM interactions. As AFM coupling is prevalent between the bipyramid units and the FM coupling within the bipyramids, it leads to the magnetic frustration. The competition between the FM and AFM interactions, and enhanced value of FM interactions in LaBaCuFeO$_5$, leads to the increase in the magnetic irreversibility between the ZFC and FC data [25]. Further, as the Fe and Cu ions can sit randomly at sites within the bipyramids it leads to the site disorder in this compound. Thus, the presence of disorder and frustration may result in the observation of magnetic cluster glass behavior in this compound. The site disorder is also possible in the La and Ba ion due to the comparative ionic radii; however, Lu has smaller ionic radii as compared to Ba. As a result, the site disorder is possibly less in the Lu based compound and the commensurate incommensurate interactions are more dominant in LuBaCuFeO$_5$ compound. Moreover, the smaller unit cell volume may have the influence on the glassy magnetic state due to the strong antiferromagnetic interactions. This may leads to the absence of glassy nature in Lu compound.

This compound shows significant change in dielectric constant with temperature along with glassy electric-dipole dynamics. Here, we would like to mention that the response in the dielectric constant in a polycrystalline compound might contain the contribution from the electrodes and from the grains/grain boundaries due to the formation of the depletion layer, with residual conductivity at high temperature. Such behavior has been observed in many other compounds [59–61]. In such cases, the imaginary part of dielectric constant is higher than the real part of dielectric at higher temperatures. However, in our compounds such behavior is not observed. Also, the reproducibility of the data on different pieces of compounds suggests the intrinsic nature of compound.

In the case of LuBaCuFeO$_5$, Cu/Fe inter-pyramidal distance decreases but the distance within the bipyramids is comparable to that of YBaCuFeO$_5$. It may also be noted that in the case of LaBaCuFeO$_5$ the distance of Cu/Fe from the apical oxygen decreases as compared to YBaCuFeO$_5$ whereas for LuBaCuFeO$_5$, this bond distance increases. The replacement of Lu in place of Y leads to decrease in inter-bipyramid distance, which strengthens the AFM interactions in comparison to the FM interaction within the bipyramids. Hence an AFM transition is observed around 175 K which is similar to that observed in YBaCuFeO$_5$ [14]. Moreover strong bifurcation in ZFC and FC around 17 K have been observed possibly due to the presence of strong spin anisotropy. The observed feature arises due to the contraction of the unit cell volume in LuBaCuFeO$_5$. A transition at 175 K arises because of the competing behavior between the AFM and FM interaction along the *c*-direction, which becomes incommensurate below the transition temperature. Since Lu has the lesser ionic radii than that of Ba, the possibility of site distortion is negligible. Also, as compared to LaBaCuFeO$_5$, the change in dielectric constant with temperature in this compound is insignificant and shows dielectric relaxation. Interestingly, the dielectric relaxation of LaBaCuFeO$_5$ is best fitted with VF law, whereas, for LuBaCuFeO$_5$, the Arrhenius law gave the best fit. The fitting of the shift in the peak temperature of the dielectric curve with VF law implies that the dipoles are strongly interacting with each other. This results in freezing of the dipoles leading to the observation of dipolar glass-like state as the temperature is lowered. In contrast, the variation of the peak temperature with frequency follows the Arrhenius law, which implies that the hopping of the dipoles due to thermal activation. This indicates that the interaction between the dipoles is not strong enough to observe the collective freezing mechanism, resulting in the absence of a glassy state. Thus, it can be inferred from the experimental data that said the lattice expansion leads the multiglass state in case of LaBaCuFeO$_5$ whereas insignificant coupling between the spin and polar degrees of freedom leads to the absence of glassy magnetic state in LuBaCuFeO$_5$.

## V. Conclusion

In our study, we have investigated the structural, magnetic and dielectric properties of rare earth layered perovskite compound LnBaCuFeO$_5$ (Ln = La, Lu). Our results on LaBaCuFeO$_5$ indicate the presence of magnetic cluster glass state at low temperature. The interesting aspect of this compound is the coexistence of magnetic cluster glass behavior along with glassy electric-dipole dynamics indicating to the presence of the multiglass behavior. In contrast, for LuBaCuFeO$_5$ AFM transition persists along with the presence of strong bifurcation at low temperature due the presence of strong spin anisotropy. In this compound the interactions between the electric dipoles is significantly weaker which results in the absence of glassy dipolar behavior. Thus, it can be said that the physical properties of these iso-electronic compounds are strongly dependent on the size of unit cell belonging to the same class. This work highlights the role of lattice structure on the magnetic and dielectric properties of layered perovskite compounds.

**Acknowledgement**

The authors acknowledge IIT Mandi for providing the experimental facilities. SL acknowledges the UGC India for SRF Fellowship**.** KM acknowledges the financial support from the CSIR project No. 03(1381)/16/EMR-II.

Table 1: Lattice parameter calculated from the Rietveld refinement of x-ray diffraction pattern of LnBaCuFeO$_5$ (Ln = La, Y, Lu).

| LnBaCuFeO$_5$ | Ln = La | Ln = Y Ref [16] | Ln = Lu |
|---|---|---|---|
| Radii (Å) | 1.032 | 0.9 | 0.861 |
| $a$ (Å) | 3.933(2) | 3.871(0) | 3.857 (2) |
| $c$ (Å) | 7.825(0) | 7.663 (1) | 7.646 (1) |
| Volume(Å$^3$) | 121.01 | 114.83 | 113.79 |
| $\chi^2$ | 1.59 | 2.18 | 1.53 |
| R-factor | 9.35 | 4.92 | 3.72 |
| RF- Factor | 7.45 | 4.18 | 4.26 |
| Inter-pyramid distance(Å) | 3.613(1) | 2.833(3) | 2.827(2) |
| Fe/Cu-O1 (Å) | 1.988 (2) | 2.117 (2) | 2.112 (3) |
| Fe-Cu distance(Å) | 3.975(4) | 3.499(2) | 3.491(4) |
| Fe-Cu distance(Å) within bi-pyramid | 3.850(4) | 4.164(2) | 4.154(1) |
| Bi-pyramid size (Å) | 4.21(2) | 4.83(0) | 4.82(1) |

**FIGURES**

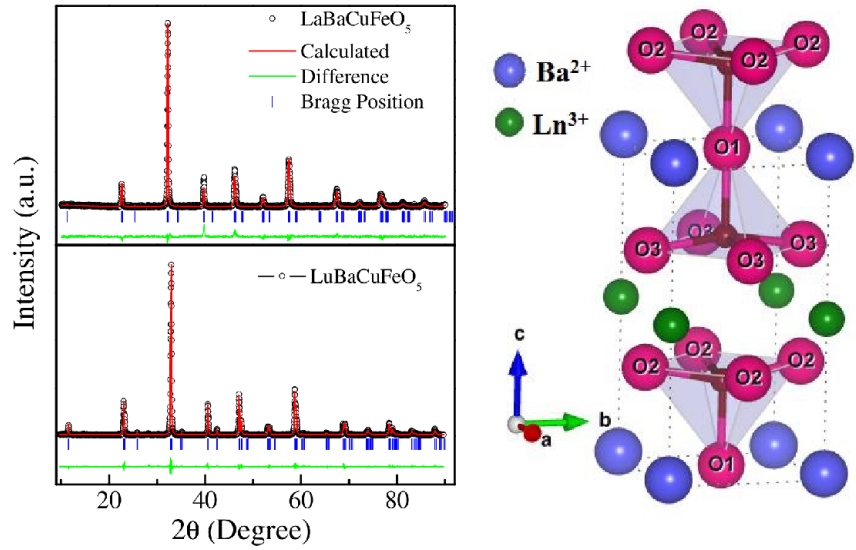

Figure 1: Rietveld refined x-ray diffraction patterns of LnBaCuFeO$_5$ (Ln = La, Lu). Right side shows the schematic of the unit cell.

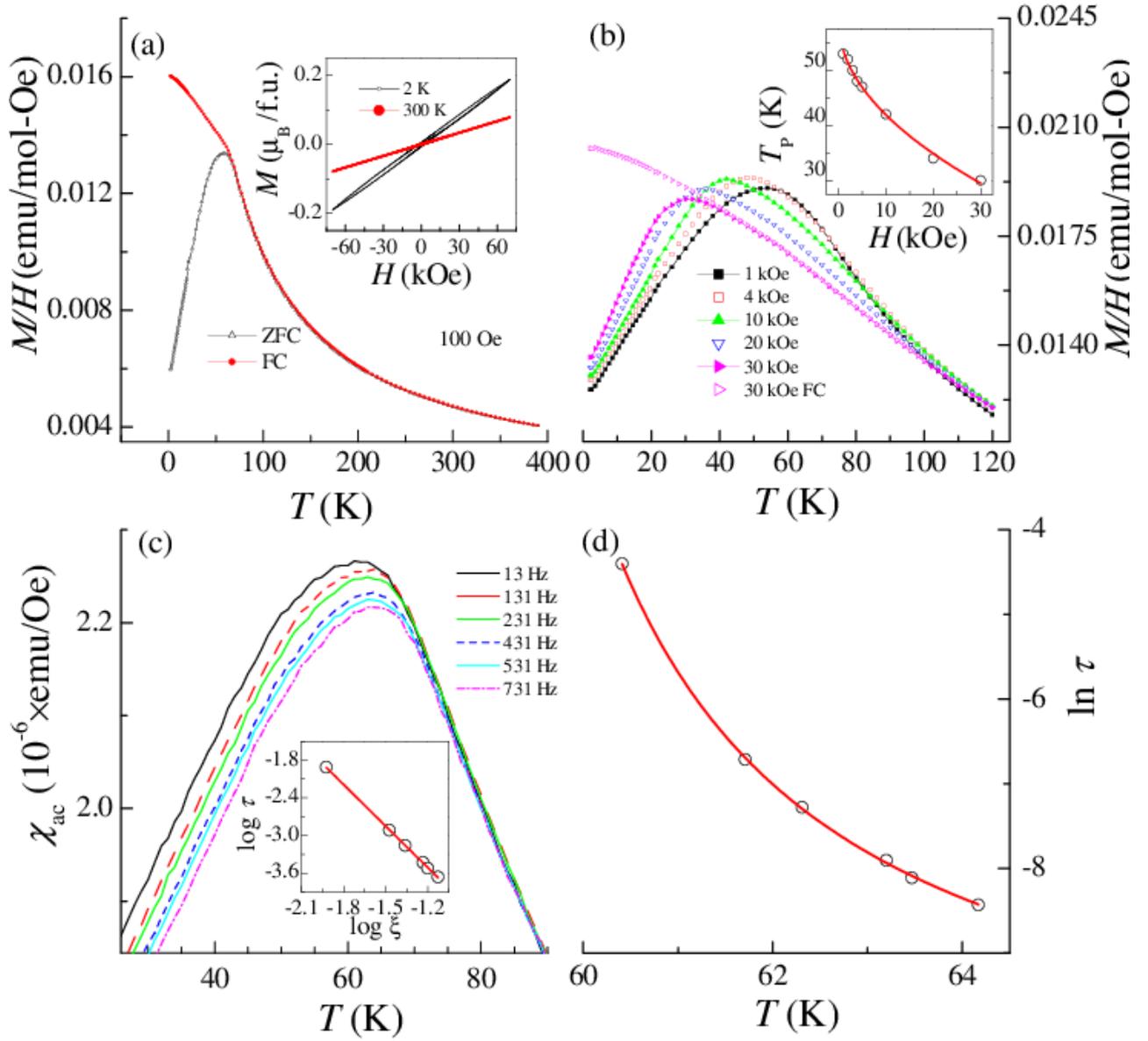

Figure 2: (a) Temperature dependence of DC susceptibility of LaBaCuFeO$_5$ obtained under ZFC and FC conditions at 100 Oe magnetic field. Inset shows $M$ vs. $H$ at 2 K and 300 K. (b) ZFC and FC magnetization data measured at different magnetic field up to 30 kOe. Inset shows the fitting of field dependence of ZFC peak temperature with A-T line. (c) The AC susceptibility measured at different frequency. Inset show the frequency dependence of freezing temperature plotted as log $\tau$ vs. log$\xi$($\xi=T_m/T_g-1$). The solid line shows the power law divergence. (d) The natural log of frequency dispersion of AC susceptibility is plotted as a function of temperature fitted with Vogel-Fulcher law.

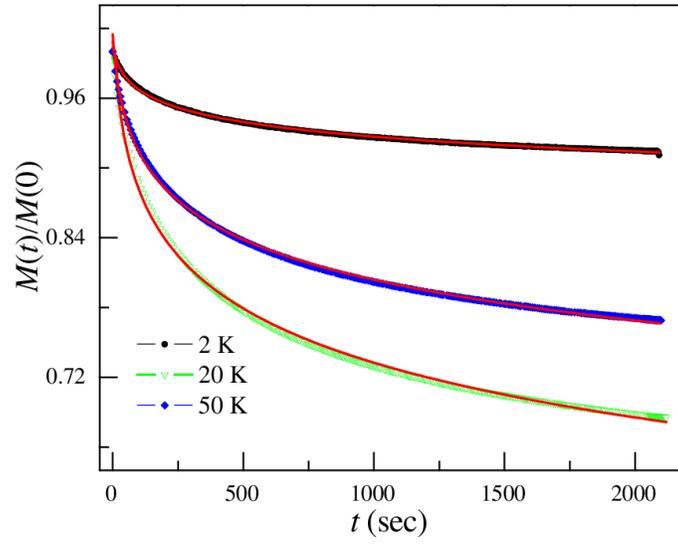

Figure 3: Relative magnetization as a function of time measured at different temperatures. Red line shows the logarithmic fit using equation (5).

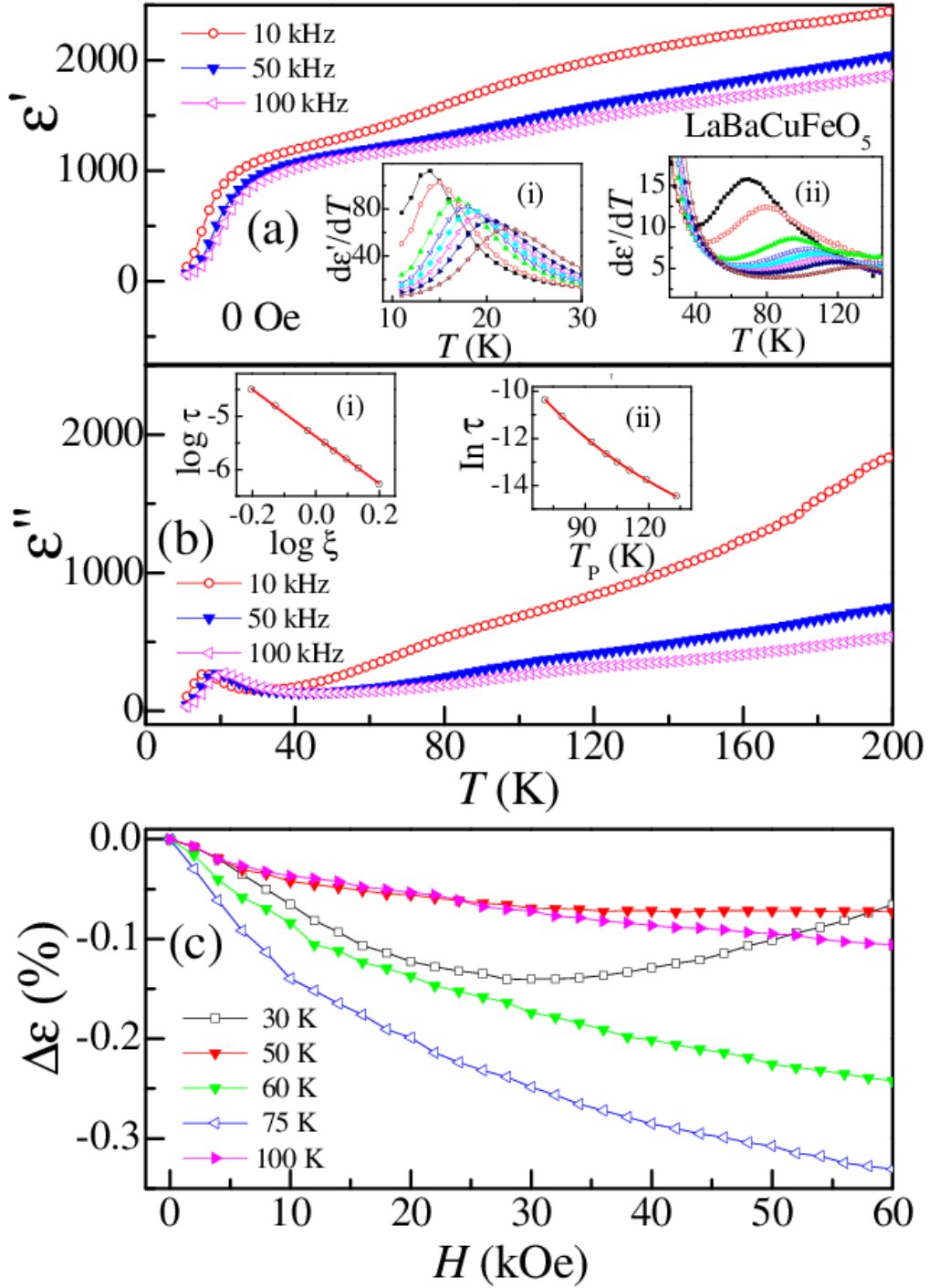

Fig 4: Temperature dependence of real part (a) and imaginary parts (b) of dielectric constant of LaBaCuFeO$_5$ measured at $f$ = 10, 50, 100 kHz. Inset of (a) (i) and (ii)) shows the derivative of real part of dielectric constant at different temperature range. Inset (b) ((i) and (ii)) shows the peak temperature ($T_P$) of derivative of dielectric constant at low temperature fitted with equation 4 and 3 respectively. (c) Relative change in the dielectric constant under the application of magnetic field measured at 10 kHz.

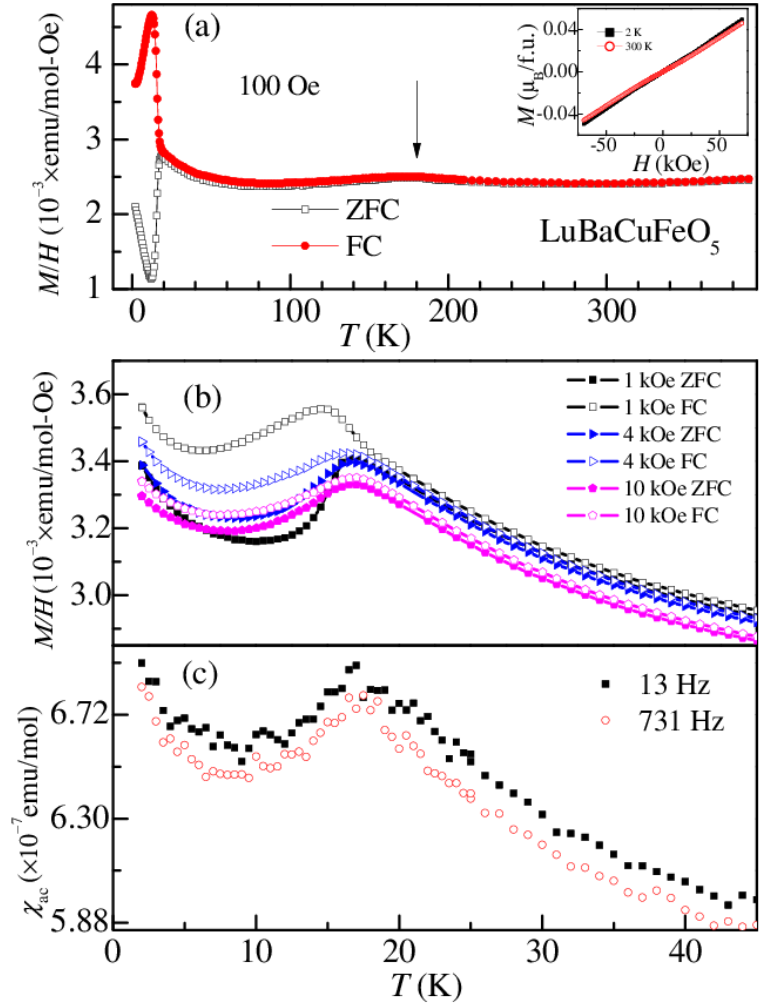

Figure 5: (a) Temperature dependence of magnetization measured at 100 Oe of field for LuBaCuFeO$_5$. Arrow points the transition temperature near about 175 K. Inset of the figure shows M versus H curves taken at 2 K and 300 K. (b) The DC magnetic susceptibility data measured at different magnetic field in ZFC and FC mode. (c) Temperature dependence of AC susceptibility measured at the different frequencies.

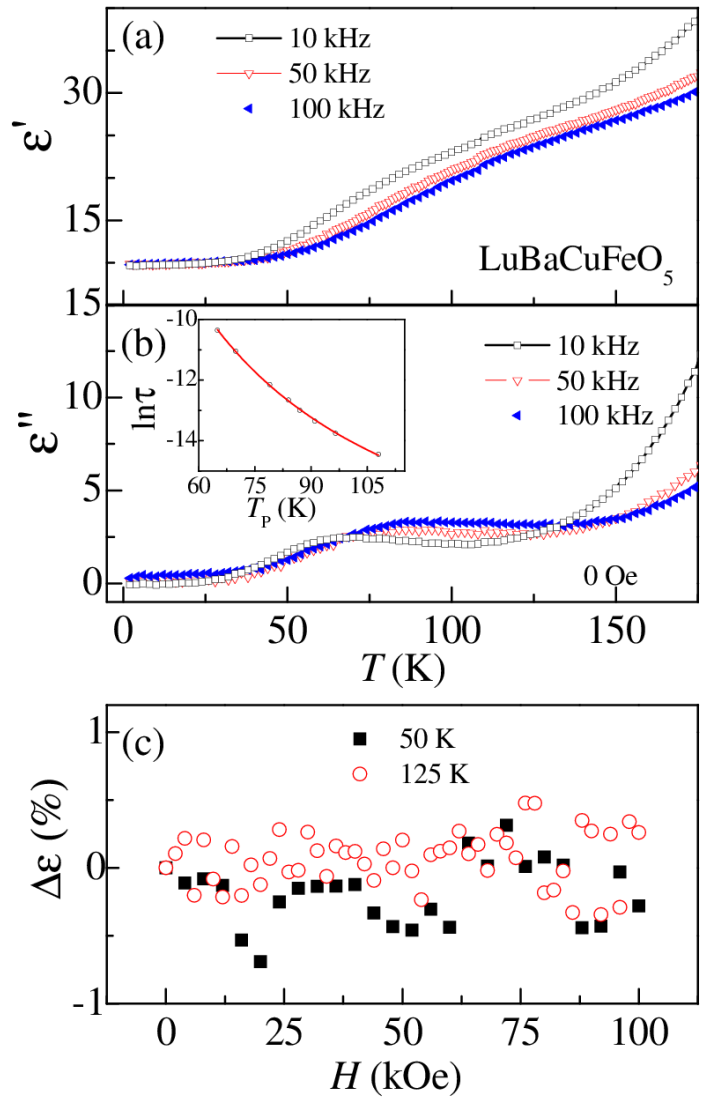

Figure 6: Temperature dependence of real part (a) and imaginary parts (b) of dielectric constant of $LuBaCuFeO_5$ measured at $f$ = 10, 50 and 100 kHz. Inset shows the Arrhenius fit of the peak temperature of the imaginary part of the dielectric constant ($\varepsilon''$) (c) Relative change in the dielectric constant under the application of magnetic field measured at 10 kHz.